\documentclass[letterpaper]{article} 
\usepackage{aaai2026}  
\usepackage{times}  
\usepackage{helvet}  
\usepackage{courier}  
\usepackage[hyphens]{url}  
\usepackage{graphicx} 
\usepackage{natbib}  
\usepackage{caption} 
\usepackage{algorithm}
\usepackage{algorithmic}

\usepackage{amsmath, amssymb}

\usepackage{newfloat}
\usepackage{listings}
\DeclareCaptionStyle{ruled}{labelfont=normalfont,labelsep=colon,strut=off} 
\floatstyle{ruled}
\newfloat{listing}{tb}{lst}{}
\floatname{listing}{Listing}
\title{Context-Aware Iterative Token Detection and Masked Transmission\\ for Wireless Token Communication}
\author {
    Junyong Shin\textsuperscript{\rm 1},
    Joohyuk Park\textsuperscript{\rm 1},
    Jihong Park\textsuperscript{\rm 2},
    Jinho Choi\textsuperscript{\rm 3},
    Yo-Seb Jeon\textsuperscript{\rm 1}
}
\affiliations {
    \textsuperscript{\rm 1}Pohang University of Science and Technology\\
     \textsuperscript{\rm 2}Singapore University of Technology and Design\\
    \textsuperscript{\rm 3}The University of Adelaide\\
    \{sjyong, joohyuk.park,  yoseb.jeon\}@postech.ac.kr, 
    jihong\_park@sutd.edu.sg,
    jinho.choi@adelaide.edu.au
}

\begin{document}

\maketitle

\begin{abstract}
The success of large-scale language models has established tokens as compact and meaningful units for natural-language representation,
which motivates token communication over wireless channels, where tokens are considered fundamental units for wireless transmission.
We propose a context-aware token communication framework that uses a pretrained masked language model (MLM) as a shared contextual probability model between the transmitter (Tx) and receiver (Rx). At Rx, we develop an iterative token detection method that jointly exploits MLM-guided contextual priors and channel observations based on a Bayesian perspective. At Tx, we additionally introduce a context-aware masking strategy which skips highly predictable token transmission to reduce transmission rate. Simulation results demonstrate that the proposed framework substantially improves reconstructed sentence quality and supports effective rate adaptation under various channel conditions.
\end{abstract}

\section{Introduction}
Recent advances in natural-language processing have demonstrated the effectiveness of processing information through discrete tokens, which represent context-aware linguistic units. As wireless communications increasingly connects devices executing token-based applications, directly transmitting tokens over the air has emerged as a relevant and promising approach, which we refer to as {\em token communication} \cite{3GPP_RAN1_122_FLS6GR, Qiao2025SemanticOrthogonality, Lee2025SemanticPacketAggregation,Oh2025HybridLanguageModel}.


Conventional communication systems treat symbols as independent entities and overlook the contextual relationships within token sequences. As a result, all tokens are transmitted with equal priority—even when some are easily predictable—and channel errors become difficult to resolve without contextual redundancy. Although recent studies have explored machine learning–based symbol detection \cite{Fan2025DEFINED}, these methods remain confined to minimizing local detection errors without leveraging inter-token dependencies. In contrast, token communication exploits contextual relationships among tokens, allowing receivers to infer missing or corrupted tokens from surrounding context. As token-based applications continue to expand, this limitation highlights the need for a shift from context-agnostic to context-aware wireless communications.

Context modeling provides a means to bridge this gap. By estimating how tokens relate to and depend on each other, context models can guide both which tokens must be transmitted and how lost information can be recovered. At the transmitter (Tx) side, context can identify predictable tokens that do not need explicit transmission, directly translating contextual redundancy into rate savings. At the receiver (Rx), context can assist in reconstructing missing or unreliable tokens from surrounding evidence, improving robustness without solely relying on stronger error protection.

Recent advances in artificial intelligence enable highly capable context models trained from large corpora of token sequences. Among them, masked language models (MLMs) such as bidirectional encoder representations from Transformers (BERT) \cite{devlin2019bert} can capture both local and long-range dependencies while inferring masked tokens based on the context. This predictive capability makes such models naturally suited for token communication: the Rx can leverage contextual knowledge as a prior, which can be combined with channel observations in a Bayesian manner. Likewise, the Tx can use the same prior to estimate which tokens are predictable and thus less critical for explicit transmission. This leads to a unified view where context models explicitly guide both transmission decisions and token inference.

Overall, we aim to address the following key question: {\em How can we jointly exploit contextual priors and channel observations to improve token transmission and reception?}
To address this, we propose a Bayesian-inspired token communication framework with rate adaptivity built on MLM-based contextual priors, consisting of:
\begin{itemize}
    \item {\bf Iterative Token Detection at Rx:} An iterative maximum a posteriori (MAP)-based token detection algorithm that refines token decisions using contextual probability and channel likelihood.

    \item {\bf Context-Aware Token Masking at Tx:} A context-aware masking strategy that suppresses predictable tokens with low entropy, reducing transmission overhead while maintaining sentence quality.
\end{itemize}
Our simulation results show clear gains in sentence quality under reduced-rate transmission. Over conventional ML detection, the proposed iterative token detection improves cosine similarity by up to 0.1769 (Europarl) and 0.1558 (WikiText-103). Additionally, under equal masking ratios, the proposed context-aware masking yields up to 0.0769 and 0.0461 improvement over random masking on the two datasets, respectively. These results highlight the effectiveness of leveraging contextual priors for both robust detection and rate adaptation.
\section{System Model}

We consider a wireless token communication system where a sequence of discrete language tokens is transmitted over a wireless fading channel. Let $\mathbf{w}$ denote the original token sequence of length $T$, given as:
\begin{equation}
\mathbf{w}=[w_1,w_2,...,w_T].
\end{equation}
To enable rate-adaptive transmission, the Tx employs a masking-based strategy that selectively masks less informative tokens according to their contextual predictability or uncertainty, producing a masked sequence 
$\mathbf{w}_{\rm m}$.
Each masked token is treated as omitted from physical transmission, and only the remaining unmasked tokens are transmitted over the channel. Hence, the masking ratio directly determines the effective transmission rate, as higher masking ratios correspond to lower physical-layer symbol usage while maintaining recoverability through contextual inference at the Rx.

Each token $w_i$ belongs to a vocabulary of size $V$ and can be represented by a binary vector of length $\lceil {\rm log}_2(V) \rceil$ through a tokenizer. Accordingly, each token element $w_{i}$ is converted into a bit sequence $\mathbf{b}_i$, defined as
\begin{equation}
\mathbf{b}_i=[b_{i,1},b_{i,2},...,b_{i,\lceil {\rm log}_2(V) \rceil}],\ \ \ b_{i,n}\in\{0,1\},\ \forall n.
\end{equation}
The resulting bit sequence $\mathbf{b}_i$ is then grouped and modulated into complex-valued symbols as
\begin{align}
    \mathbf{s}_i=[s_{i,1},s_{i,2},...,s_{i,\left\lceil\lceil {\rm log}_2(V) \rceil/m\right\rceil} ],\ \ \ s_{i,k}\in\mathcal{S},\ \forall k,
\end{align}
where each symbol $s_{i,k}$ represents a group of $m$ consecutive bits that are jointly mapped onto a complex symbol constellation $\mathcal{S}$.
For example, a $2^m$-ary Quadrature Amplitude Modulation (QAM) scheme can be employed for symbol mapping,
which is widely used in digital communication systems.
The modulated symbols are transmitted over a Rayleigh block-fading channel \cite{goldsmith2005wireless, Shin2025ESC_MVQ, Kang2025MIMOBoosting} modeled as
\begin{align}
    {\bf y}_i=h\sqrt{p_{\rm tx}}{\bf s}_i+{\bf n}_i,
\end{align}
where $h\in \mathbb{C}$ denotes the complex fading coefficient that remains constant over one transmission block, $p_{\rm tx}$ is the transmit power, and $\mathbf{n}_i\sim \mathcal{CN}(0,\sigma^2\mathbf{I})$ is additive white Gaussian noise (AWGN). The signal-to-noise ratio (SNR) in dB is defined as
\begin{align}
    \text{SNR (dB)}=10~{\rm log}_{10}\left(\frac{p_{\rm tx}\mathbb{E}[|h|^2]}{\sigma^2}\right).
\end{align}


Given the entire observation ${\bf y}=[\mathbf{y}_1,...,\mathbf{y}_T]$, the Rx computes the symbol-wise likelihood as
\begin{align}
    P(y_{i,k}|\hat{s}_{i,k})=\frac{1}{\pi\sigma^2}{\rm exp}\left(-\frac{|y_{i,k}-h\sqrt{p_{\rm tx}}\hat{s}_{i,k}|^2}{\sigma^2}\right),\ \ \nonumber \\ \hat{s}_{i,k}\in\mathcal{S},\ \forall i,k.
\end{align}
From these, the token-level likelihoods $P({\bf y}_i|w_i)$ are derived as
\begin{align}
    P(\mathbf{y}_i|w_i)=\prod_{k=1}^{\left\lceil\lceil {\rm log}_2(V) \rceil/m\right\rceil}P(y_{i,k}|\hat{s}_{i,k}).
\end{align}
This probabilistic representation $P({\bf y}_i|w_i)$ expresses the likelihood of receiving token $w_i$ under channel distortion.
Because no channel information is conveyed for masked positions, their likelihoods are set as a uniform distribution over the vocabulary, i.e., $P({\bf y}_{i^*}|w_{i^*})=1/V$ for any masked token position $i^*$.

The detected token sequence is expressed as
\begin{align}
    \mathbf{\hat w}=[{\hat w}_1, {\hat w}_2,..., {\hat w}_T].
\end{align}
If a conventional wireless communication system were applied, token detection would rely solely on maximizing this channel likelihood, following the maximum likelihood (ML) detection principle as
\begin{align}
    \hat{w}_i=\underset{w_i}{{\rm argmax}}\ P(\mathbf{y}_i|w_i).
\end{align}
However, this approach ignores the contextual dependencies among tokens—each token is treated independently, even when certain tokens are highly predictable given their neighbors. As a result, ML detection becomes suboptimal in wireless token communication, where contextual information can play a critical role in recovering omitted or corrupted tokens.

\begin{figure*}[t]
\centering
\includegraphics[width=1.8\columnwidth]{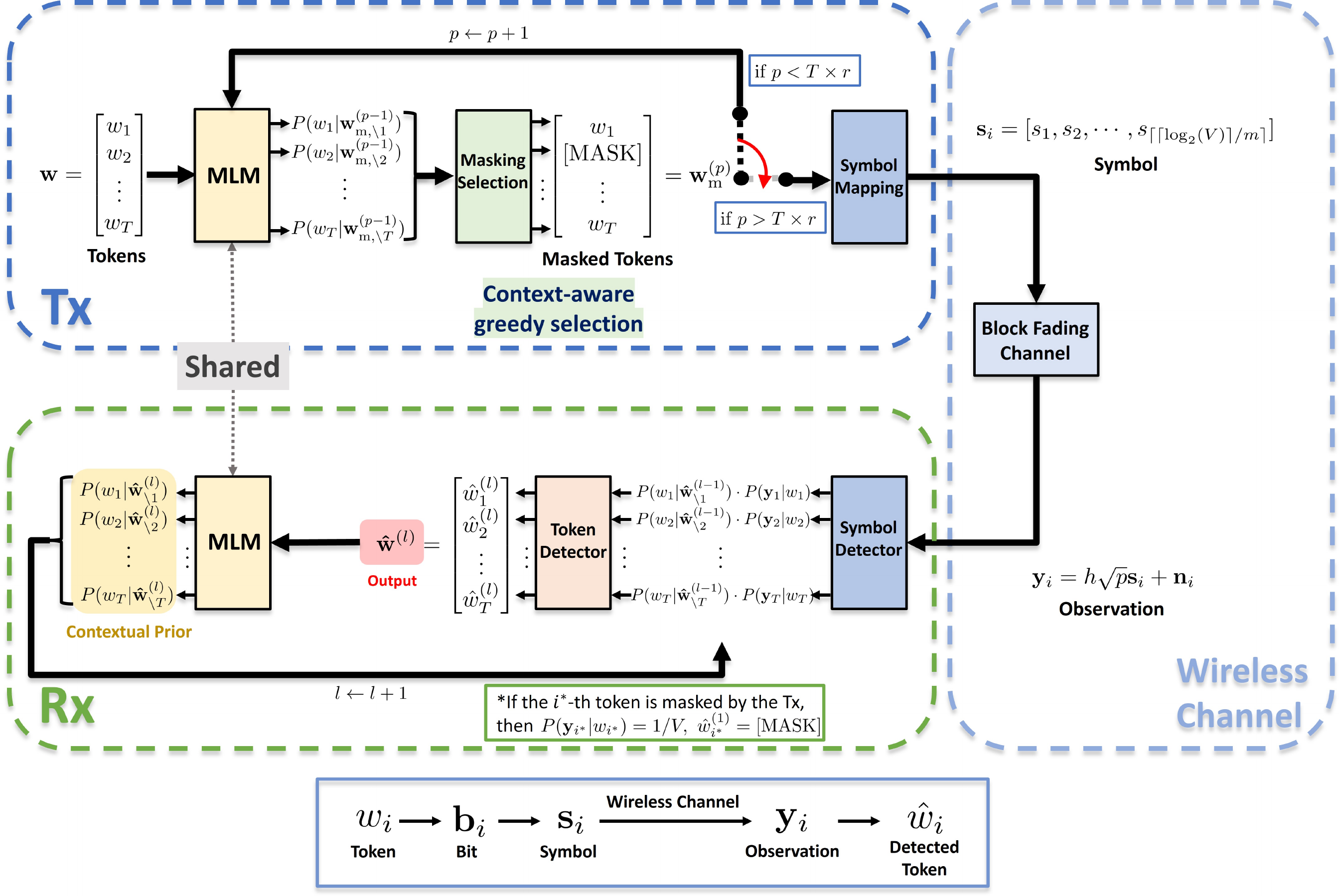}
\caption{An illustration of the proposed token communication framework.}
\label{fig1}
\end{figure*}

\section{Iterative Token Detection for Context-Aware Wireless Token Communication}
To overcome the limitations of conventional ML detection, which relies solely on channel observations and ignores contextual dependencies among tokens, we reformulate token detection as a MAP problem. This formulation enables the Rx to jointly consider both the channel likelihood and a {\bf contextual prior} that captures inter-token dependencies. The contextual prior represents the conditional probability of a token given its surrounding tokens, serving as a probabilistic reference that guides detection when channel information is unreliable or missing.

To compute this contextual prior, we employ a MLM, which estimates token-level conditional probabilities in masked position by leveraging contextual information from the surrounding tokens. Leveraging this shared prior knowledge, the proposed framework establishes a unified design for both Tx and Rx: the Rx iteratively refines uncertain token estimates using MLM-derived contextual priors, while the Tx exploits the same priors in reverse to mask predictable tokens and thereby achieve effective rate adaptation. The overall proposed framework is illustrated in Fig. \ref{fig1}.


\subsection{MLM as Contextual Token Prior Model}\label{Sec: contextual prior}
In our token communication framework, MLM is adopted as a contextual token prior model shared by both the Tx and Rx.
The MLM produces a contextual probability distribution over the vocabulary for any masked token position, leveraging surrounding tokens to infer the most plausible candidate.
Formally, given an input sequence of tokens $\mathbf{x}=[x_1,...,x_T]$, MLM estimates a conditional categorical distribution over the vocabulary at each position:
\begin{align}
    P(x_i|\mathbf{x}_{\backslash i}) =[{\rm MLM}(\mathbf{x}_{\backslash i})]_i,
\end{align}
where ${\rm MLM}(\cdot)$ represents the MLM function, and $\mathbf{x}_{\backslash i}\triangleq [x_1,...,x_{i-1},x_{i+1},...,x_T]$ denotes the sequence with $i$-th element ommited. This operation intrinsically models token dependency and contextual semantics.

Let $[{\rm MASK}]$ denote the special mask token used by MLM to hide selected positions. For any masking set $\mathcal{M}\subset \{1,...,T\}$, we define a masking operator:
\begin{align}
    [{\rm Mask}(\mathbf{w};\mathcal{M})]_i=
    \begin{cases}
        [{\rm MASK}],\ &i\in \mathcal{M},\\
        w_i, &i \notin \mathcal{M},
    \end{cases}
\end{align}
and express the masked token sequence used for inference as $\mathbf{w}_{\rm m} = {\rm Mask}(\mathbf{w};\mathcal{M})$. If $\mathcal{M}=\varnothing$, this operation reduces to the original sequence, i.e., $\mathbf{w}=\mathbf{w}_{\rm m} = {\rm Mask}(\mathbf{w};\mathcal{M})$.
For a specific index $i$, the conditional distribution of the true token $w_i$ is computed by additionally masking the $i$-th position of the masked token sequence $\mathbf{w}_{\rm m}$, yielding $\mathbf{w}_{{\rm m},\backslash i}$, and feeding it into the MLM as input:
replacing only the $i$-th position with $[{\rm MASK}]$:
\begin{align}\label{eq: mask BERT2}
    P(w_i|\mathbf{w}_{{\rm m},\backslash i}) =[{\rm MLM}(\mathbf{w}_{{\rm m},\backslash i})]_i,
\end{align}
which provides a prior belief about $w_i$ given the observed or unmasked context tokens. We refer to this prior belief as the \textbf{contextual prior}, as it captures the token-level probability distribution conditioned on the surrounding context.

In the proposed wireless token communication system, this contextual prior plays two complementary roles:
\begin{itemize}
    \item {\bf Tx-side:} The contextual prior $P(w_i|\mathbf{w}_{{\rm m},\backslash i})$ provides an entropy metric to determine which tokens are more redundant and can be masked for rate adaptation.

    \item {\bf Rx-side:} The same contextual prior $P(w_i|\mathbf{\hat w}^{(l-1)}_{\backslash i})$ serves as the probabilistic reference during iterative detection, refining token estimates by combining contextual knowledge with channel observations.
\end{itemize}
Thus, MLM establishes a shared contextual probabilistic model between Tx and Rx, enabling semantic-aware encoding and robust inference over wireless channels.

\subsection{Rx Strategy: Context-Aware Iterative Token Detection}
This part builds on the contextual prior established in the last section to implement Bayesian token detection.
By iteratively refining token estimates using contextual priors derived from MLM, the Rx progressively approaches the MAP token detection.

At the Rx, the channel observation is first converted into a set of token-wise likelihoods representing the equivalent received token sequence $\mathbf{y}$. The objective of iterative token detection is to solve a MAP token detection problem:
\begin{equation}
\hat{w}_i=\underset{w_i}{{\rm argmax}}\ P(w_i|\mathbf{y}).
\end{equation}
Applying Bayes' rule and using the total probability theorem over $\mathbf{w}_{\backslash i}$ yields
\begin{align}
\hat{w}_i&=\underset{w_i}{\text{argmax}}\ P(\mathbf{y}|w_i)P(w_i)\\
&=\underset{w_i}{\text{argmax}}\ \sum_{\mathbf{w}_{\backslash i}}P(\mathbf{y}|\mathbf{w})P(\mathbf{w}).
\end{align}
Moreover, using the token-conditional independence of observations, the MAP rules becomes
\begin{align}
\hat{w}_i&=\underset{w_i}{\text{argmax}}\ \sum_{\mathbf{w}_{\backslash i}}\prod^T_{j=1}P({\bf y}_j|w_j) P(\mathbf{w})
\\&=\underset{w_i}{\text{argmax}}\ P({\bf y}_i|w_i) \sum_{\mathbf{w}_{\backslash i}}\prod_{j\neq i}P({\bf y}_j|w_j) 
  P(\mathbf{w})
\\&=\underset{w_i}{\text{argmax}}\ P({\bf y}_i|w_i) \sum_{\mathbf{w}_{\backslash i}} 
  \prod_{j\neq i}P({\bf y}_j|w_j)
P(w_i|\mathbf{w}_{\backslash i})P(\mathbf{w}_{\backslash i}).
\end{align}
To avoid the entanglement and intractability arising from marginalizing over all combinations of $\mathbf{w}_{\backslash i}$ — since each token’s probability is inherently coupled with others in the sequence, forming probabilistic loops — we adopt a widely used single-sequence approximation in iterative detection and decoding frameworks \cite{Tuchler2002TurboEqualization}.
The key insight is that the posterior probability mass $P(\mathbf{w}_{\backslash i})$ tends to be highly concentrated around the most likely token sequence estimate from the previous iteration. Therefore, we approximate
\begin{align}
    P(\mathbf{w}_{\backslash i})\approx
    \begin{cases}
        1,\ \ \ \text{if }\mathbf{w}_{\backslash i}=\mathbf{\hat w}_{\backslash i}^{(l-1)},\\
        0,\ \ \ \text{otherwise,}
    \end{cases}
\end{align}
which allows the summation over $\mathbf{w}_{\backslash i}$ to be replaced by the single detected sequence from iteration $l-1$. This leads to the following tractable MAP detection rule:
\begin{equation}\label{eq: iteration}
\hat{w}_i^{(l)}= \underset{w_i}{\text{argmax}}\ P({\bf y}_i|w_i)P(w_i|\mathbf{\hat w}_{\backslash i}^{(l-1)}).
\end{equation}

Here, the first term $P(\mathbf{y}_i|w_i)$ corresponds to the channel likelihood introduced in the previous section, while the second term $P(w_i|\mathbf{\hat w}_{\backslash i }^{(l-1)})$ represents the contextual prior derived from MLM.
Unlike conventional ML detection, which relies solely on $P(\mathbf{y}_i|w_i)$, the proposed MAP formulation jointly considers both the channel likelihood and the context-based prior that captures inter-token dependencies.
This integration enables more reliable token inference, particularly in cases where the received symbols are corrupted or when certain tokens are omitted by the Tx, since contextual knowledge can compensate for incomplete or noisy channel observations.

For the first iteration ($l=1$), no contextual prior information is available at the Rx. Therefore, the prior distribution is assumed to be uniform over the entire vocabulary:
\begin{align}
P(w_i|\mathbf{\hat w}_{\backslash i}^{(0)})=\frac{1}{V},\ \forall i.
\end{align}
If the $i^*$-th token is masked by the Tx, the corresponding received token at the first iteration is initialized as the mask token, i.e.,
\begin{align}
    \hat{w}_{i^*}^{(1)} = [{\rm MASK}], \forall i^*\in\mathcal{M},
\end{align}
where $\mathcal{M}$ denotes the index set of tokens masked at the Tx.
Since these masked tokens are not physically transmitted over the wireless channel, the Rx has no valid channel observation associated with it. Consequently, the likelihood of the received signal for masked tokens is treated as unbiased and expressed as
\begin{align}
P({\bf y}_{i^*}|w_{i^*})=\frac{1}{V},\ \forall i^*\in\mathcal{M}.
\end{align}
This initialization ensures that all masked positions begin with uniform uncertainty, allowing the iterative detection process to progressively refine their estimates using both contextual and channel-based information.
For $l\geq 2$, priors are refined via the shared MLM.
Specifically, the Rx masks position $i$ in $\mathbf{\hat w}^{(l)}$, thereby constructing $\mathbf{\hat w}_{\backslash i}^{(l)}$, and feeds it as MLM input:
\begin{align}
P(w_i|\mathbf{\hat w}_{\backslash i}^{(l)})=[{\rm MLM}(\mathbf{\hat w}_{\backslash i}^{(l)})]_i,
\end{align}
which is then used as the prior for the next iteration. This iterative process continues until a maximum iteration count, gradually improving its token prior and detection output.

\subsection{Tx Strategy: Context-Aware Masking for Rate-Adaptive Token Transmission}
The Tx side complements the iterative detection by exploiting the same contextual prior in reverse: tokens that exhibit low uncertainty under the MLM are masked and not physically transmitted.
This mechanism enables effective rate adaptation, as predictable tokens rely on contextual inference at the Rx, while only uncertain tokens are explicitly transmitted.

To formalize this process, the Tx quantifies the transmission rate by the masking ratio 
$r\in[0,1]$, which determines how many tokens are omitted from physical transmission.
The Tx exploits the shared MLM to decide which tokens can be reliably inferred by the Rx—thus enabling rate adaptation under a given transmission budget.
Let $r$ denote the masking ratio, resulting in a masking budget of $\lfloor Tr\rfloor$.

We identify the masking set $\mathcal{M}^{(p)}\subset \{1,...,T\}$ through a greedy selection process performed over $p$ iterations. At iteration  $p\in\{1,...,\lfloor Tr \rfloor\}$, the current masked sequence is denoted as $\mathbf{w}_{\rm m}^{(p-1)}$. 
Given $\mathbf{w}^{(p-1)}_{\rm m}$, the Tx evaluates, for every unmasked position $i \notin \mathcal{M}^{(p-1)}$, the MLM predictive distribution obtained as
\begin{align}
P(w_i|\mathbf{w}^{(p-1)}_{{\rm m},\backslash i})=[{\rm MLM}(\mathbf{w}^{(p-1)}_{{\rm m},\backslash i})]_i.
\end{align}
The predictability of position $i$ is quantified by its token-wise entropy
\begin{align}
H_i^{(p)}=-\sum_v P(w_i=v|\mathbf{w}^{(p-1)}_{{\rm m},\backslash i}){\rm log}_2P(w_i=v|\mathbf{w}^{(p-1)}_{{\rm m},\backslash i}),
\end{align}
where a lower entropy implies that the token is easier for the Rx to estimate without explicit transmission.
Utilizing this formulation, the Tx selects the most predictable token index to be masked next:
\begin{align}
i^*_{p} = \underset{i\notin \mathcal{M}^{(p-1)}}{\text{argmin}}\ H_i^{(p)}.
\end{align}
Accordingly, the masking set and masked sequence are updated as
$\mathcal{M}^{(p)}=\mathcal{M}^{(p-1)}\cup\{i^*_p\}$, $\mathbf{w}_{\rm m}^{(p)}={\rm Mask}(\mathbf{w};\mathcal{M}^{(p)})$, and this procedure iterates until $p>Tr$. By construction, $\mathbf{w}^{(0)}_{\rm m}=\mathbf{w}$ establishes.

Since each transmitted (i.e., unmasked) token requires $\lceil {\rm log}_2(V) \rceil$ bits, masking reduces the physical-layer transmission rate while relying on the shared language model to restore the missing semantics at the Rx. This tightly couples the communication rate with the linguistic predictability of the source content.
\section{Simulation Results}

We evaluate the proposed context-aware token communication framework using a text transmission task.
The Europarl corpus \cite{koehn2005europarl} and the WikiText-103 dataset \cite{merity2016pointer} are
used as the source text dataset, and tokens are generated by WordPiece \cite{wu2016google}-based tokenization. The BERT model in \cite{devlin2019bert} is employed as the shared MLM at both the Tx and Rx. We consider packets consisting of $T=128$ tokens, where each token is represented by 15 bits. For physical-layer transmission, 16-QAM modulation is applied over a Rayleigh block fading channel.
To assess the semantic correctness of the received text, we use a cosine similarity metric, denoted as {\em SIM}, between sentence embeddings obtained from the “all-MiniLM-L6-v2” SentenceTransformer model \cite{reimers2019sentencebert}. This embedding-based similarity effectively measures meaning preservation, which aligns with the goal of token communication.

\begin{figure}[t]
\centering
\begin{minipage}{1\columnwidth}
    \centering
    \includegraphics[width=0.9\linewidth]{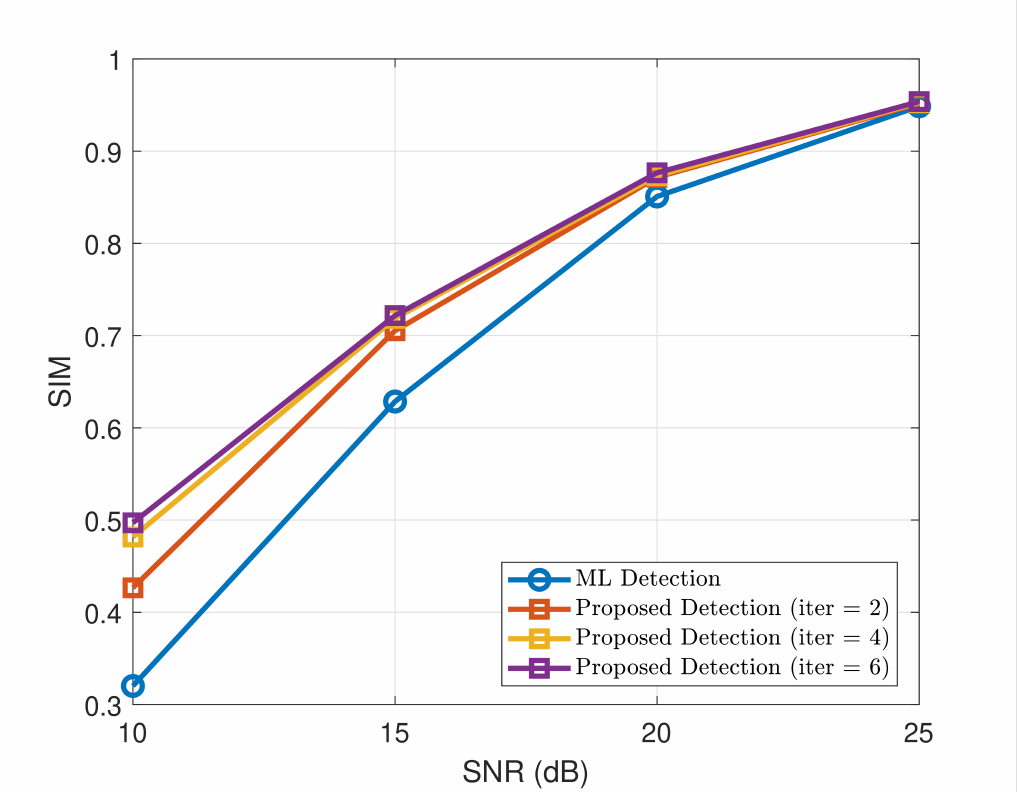}
    \caption*{(a) The Europarl corpus dataset}
    \label{fig:sub_a}
\end{minipage}
\hfill
\begin{minipage}{1\columnwidth}
    \centering
    \includegraphics[width=0.9\linewidth]{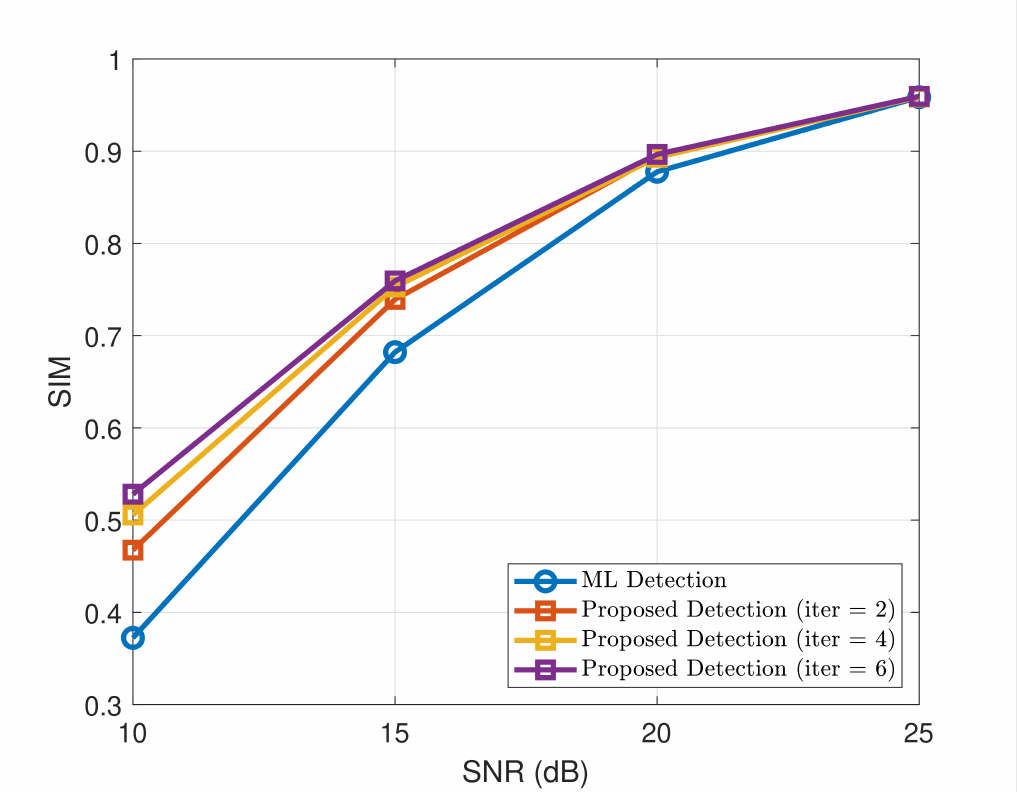}
    \caption*{(b) The WikiText-103 dataset}
    \label{fig:sub_b}
\end{minipage}
\caption{Performance of the proposed iterative token detection strategy (Rx-side only) under different iteration counts.} \label{fig2}
\end{figure}

Fig. \ref{fig2} illustrates the performance of the proposed iterative token detection strategy without applying the Tx masking scheme (i.e., full token transmission). The baseline {\em ML Detection} performs token estimation solely based on the channel likelihood $P({\bf y}_i|w_i)$, without incorporating the contextual prior $P(w_i|\hat{\mathbf{w}}^{(l-1)}_{\backslash i})$ derived from MLM. In contrast, the {\em Proposed Detection} incorporates contextual priors from MLM and iteratively refines the token estimates using eq. \eqref{eq: iteration}.
Fig. \ref{fig2} shows that even a single iteration significantly improves performance over the baseline, particularly in low-SNR regimes. As the number of iterations increases from ${\rm iter}=2$ to ${\rm iter}=6$, the SIM value improves steadily and nearly converges at five iterations. Overall, the proposed Rx strategy clearly outperforms the baseline, indicating that semantic-aware iterative refinement can effectively compensate for channel-induced distortion.

\begin{figure}[t]
\centering
\begin{minipage}{1\columnwidth}
    \centering
    \includegraphics[width=0.9\linewidth]{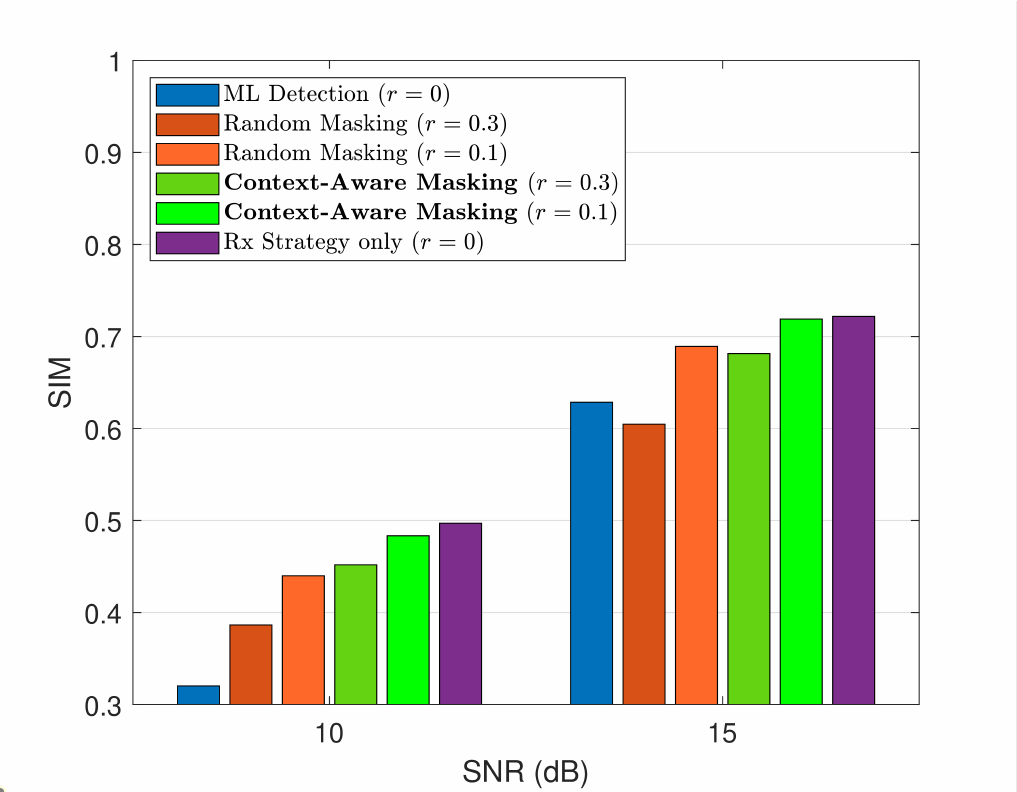}
    \caption*{(a) The Europarl corpus dataset}
    \label{fig:sub_a}
\end{minipage}
\hfill
\begin{minipage}{1\columnwidth}
    \centering
    \includegraphics[width=0.9\linewidth]{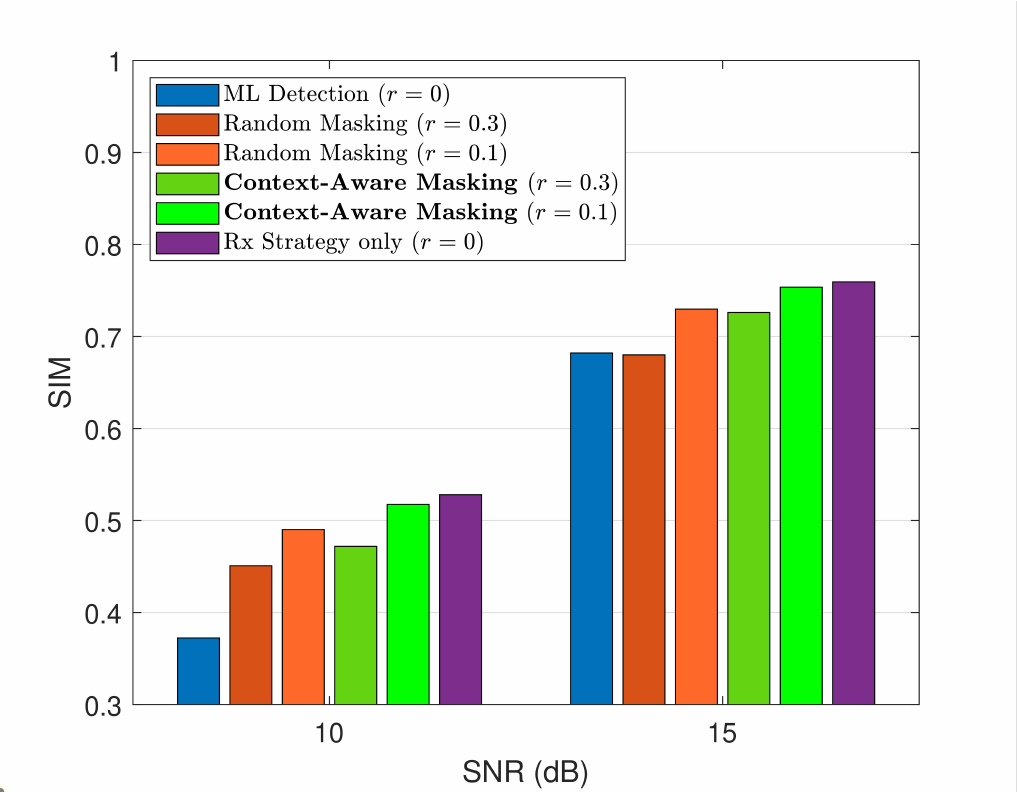}
    \caption*{(b) The WikiText-103 dataset}
    \label{fig:sub_b}
\end{minipage}
\caption{Performance comparison of the joint Tx–Rx strategy with different masking ratios and masking policies.} \label{fig3}
\end{figure}

Fig. \ref{fig3} presents the performance when the Tx adopts the proposed context-aware masking strategy to reduce the transmission rate. Except for the {\em ML Detection}, the proposed iterative token detection method is adopted, and the number of iterations $\text{iter}=6$ and masking ratios of $r\in\{0.1,0.3\}$ are adopted. Aditionally, as Tx strategies, two masking policies are compared:
\begin{itemize}
\item{\bf Random Masking}: the Tx removes tokens at arbitrary positions without considering the predictability of Rx, serving as a lower-bound inference.
\item{\bf Context-Aware Masking}: the proposed greedy masking strategy that selectively masks highly predictable tokens based on MLM inference.
\end{itemize}
As shown in the figure, random masking causes noticeable performance degradation due to loss of important semantic information. In contrast, the proposed context-aware masking achieves significantly higher SIM levels than that of random masking, for each masking ratios. Particularly, when $r=0.1$, the performance of proposed context-aware masking remains very close to the {\em Rx Strategy only} (without masking) case, demonstrating strong rate-adaptation capability with minimal degradation in context quality. These results confirm that the proposed joint Tx–Rx strategy successfully harnesses context redundancy to reduce physical-layer transmission cost while maintaining high-quality text reconstruction at the Rx.

\section{Conclusion}
We presented a Bayesian-inspired token communication framework that integrates MLM as a contextual prior model. In this framework, the Tx selectively masks predictable tokens to reduce rate, while the Rx performs iterative MAP-based detection using contextual priors and channel likelihoods. Experiments demonstrate significant improvements in reconstructed sentence quality and strong rate-adaptive behavior under noisy channels. While this study focuses on text token transmission, extending the framework to multimodal token scenarios could be an interesting topic for future research.

\section{Acknowledgments}
This work was supported in part by the National Research Foundation of Korea (NRF) grant funded by the Korea government (MSIT) (No. RS-2024-00453301, 50\%), in part by the Institute of Information \& Communications Technology Planning \& Evaluation (IITP)-ITRC (Information Technology Research Center) grant funded by MSIT (IITP-2025-RS-2023-00259991, 25\%), and in part by SUTD Kickstarter Initiative
(SKI 2021 06 08, 25\%).

\bigskip
\bibliography{references}

\end{document}